\documentclass[12pt]{iopart}

\usepackage{color}
\usepackage{xcolor}
\usepackage{graphicx}
\usepackage{dcolumn}
\usepackage{setstack}
\usepackage{amsthm}
\usepackage{array}
\usepackage[breaklinks]{hyperref}
\usepackage[all]{hypcap}
\usepackage[latin10]{inputenc}
\usepackage{amsfonts}
\usepackage[english]{babel}
\usepackage{epstopdf}
\newcommand{\comillas}[1]{\textquotedblleft #1\textquotedblright}

\newcounter{examp}
\begin{document}
\title{Exploiting non-trivial spatio-temporal correlations of thermal radiation for sunlight harvesting}

\author{A. M. De Mendoza$^1$, F. Caycedo-Soler$^2$, P. Manrique$^3$, L. Quiroga$^1$, F. J. Rodriguez$^1$, and N F. Johnson$^3$.}

\address{$^1$ Physics Department, Universidad de Los Andes, Cra 1 Nș 18A- 12 Bogot\'a, Colombia.}
\address{$^2$ Institut f\"ur Theoretische Physik, Universit\"at Ulm, Albert-Einstein-Allee 11, 89073 Ulm, Germany.}
\address{$^3$ Physics Department, University of Miami, Coral Gables, FL  33124 Miami, USA.}

\begin{abstract}{


The promise of any small improvement in the performance of light-harvesting devices, is sufficient to drive enormous experimental efforts. However these efforts are almost exclusively focused on enhancing the power conversion efficiency with specific material properties and harvesting layers thickness, without exploiting the correlations present in sunlight -- in part because such correlations are assumed to have negligible effect. Here we show, by contrast, that these spatio-temporal correlations are sufficiently relevant that the use of specific detector geometries would significantly improve the performance of harvesting devices. The resulting increase in the absorption efficiency, as the primary step of energy conversion, may also act as a potential driving mechanism for artificial photosynthetic systems. Our analysis presents design guidelines for optimal detector geometries with realistic incident intensities based on current technological capabilities.}
\end{abstract}

\noindent{\it Thermal light, light correlations, harvesting enhancement\/}\\
\submitto{\jpb}
\maketitle
	
\section{Introduction}	

The  discovery of the photoelectric effect  \cite{Hertz,Einstein} initiated the development of sunlight conversion, which, by mimicking nature's capability to utilize this abundant energy source, is a promising alternative for energy production. The existing photovoltaic technologies have to date, a maximum conversion efficiency between 10\% and 46\%\footnote{The solar conversion efficiency is usually defined for photovoltaic systems as electrical power output (W/cm$^{2}$) divided by incident solar irradiance (W/cm$^{2}$) measured over the entire solar spectrum \cite{Blankenship,Nrel}. This definition has been only recently set on equal footing for photosynthetic conversion \cite{Blankenship}.}, acquired with relatively expensive materials\cite{Blankenship,Nrel,Kotter,SCreview}. The designs encompass the competing elements of low excitonic recombination  and spectrally broad high harvesting cross section. More economical technological alternatives have been proposed, e.g., nano-antennas technology, which for solar frequencies, succeed in conversion of absorbed electromagnetic radiation with up to 80\%-90\% efficiency in a primary collection, followed however by a rectifying stage (to convert AC into DC currents), where efficiencies drop down to 0.01\% \cite{Vandenbosch}. The continuous improvement of this latter stage \cite{Sharma,SAreview}, still set nano-antennas as good candidates for competitive energy production, as well. Both solar cells and nano-antennae would benefit of an increased absorption cross section \cite{SAreview,Mashaal1}. The improvement that has been achieved so far was based on the development of material technologies, while the potential coming from exploiting correlations present in the absorbed photons has been overlooked. These correlations have lately found renewed attention, as illustrated by the  violation of Bell inequalities or the observation of the Hanbury-Brown-Twiss effect with natural sunlight\cite{Adesso}. In fact, recent research has shown a favorable potential for thermal light spatio-temporal correlations to assist natural light harvesting in bacterial photosynthesis \cite{Manrique}. Therefore, a fundamental open question concerns the extent to which such correlations might be exploited in order to improve artificial light harvesting for energy conversion technologies\cite{sim,brumer1}. \\

In this paper we prove that the detection geometries can be optimized for the specific radiation features, and we particularly show how thermal radiation correlations can be exploited to improve the capture of abundant energy provided by the Sun. Based on well tested theoretical results, our findings give useful directions to lead a more efficient design of solar energy collection devices, which illustrates the possibility to enhance light harvesting by engineering the receptors-architecture when spatio-temporal correlations present in the thermal radiation are exploited.

\section{Methods: factorial moments generating function} \label{method}

Photodetection can be well characterized by the statistical properties of the incident light, and the triggering of photo-electric events \cite{SAreview}. All the statistical information of the stochastic  photodetection events  is contained in its generating function, which serves to cast the \textit{factorial moments generating function}

 \begin{equation}
G(\left\{s_i\right\})=\prod_{j=1}^T \prod^{N}_{k=1} (1+\varpi_j b_k s_i)^{-1},
\label{G_gen}
\end{equation}

\noindent where $\{s_i\}$ is a set of expansion parameters, $N$ is the number of detectors of the photodetection system and $T$ is the time interval during which each detector is open to detect light -- called \textit{detection time} or \textit{counting time} \cite{Bures71}. The detection time value defines the capability of the detection system to perceive the light's temporal coherence. 
Particularly, in light harvesting devices it depends on the specific detection mechanism and so do the results we subsequently present. In the last expression $\varpi_j$ and $b_k$ are the eigenvalues of the temporal and spatial Fredholm equations, $\int^T_0 dt_2 \Gamma(t_1-t_2) \psi_j(t_2)=\varpi_j\psi_j(t_1)$  and $\sum_{l=1}^N \sqrt{\alpha_k} \Gamma^*_{k,l}\sqrt{\alpha_l}\phi_l(t)=b_k \phi_k(t)$ which depend on the correlation functions of the electric field at different positions, i.e., the spatial correlations $\Gamma_{k,l}=\langle \hat{\vec{E}}(\vec{r}_k,t) \hat{\vec{E}}(\vec{r}_l,t)\rangle$, and on different times, i.e., the temporal correlations $\Gamma(t_1,t_2)=\langle\hat{\vec{E}}(\vec{r},t_1) \hat{\vec{E}}(\vec{r},t_2)\rangle$. In particular, the normalized correlation functions 

\begin{eqnarray}\label{fcoh}
\gamma_{k,l}=\frac{\Gamma_{k,l}}{\Gamma_{k,k}}=2\frac{J(\nu_{k,l})}{\nu_{k,l}},\\
\gamma(t_1,t_2)=\frac{\Gamma(t_1,t_2)}{\Gamma(t_1,t_1)}=\frac{\sin \left[\frac{2\pi(t_1-t_2)}{\tau_c}\right]}{\left(\frac{2\pi(t_1-t_2)}{\tau_c}\right)}
\end{eqnarray}	

\noindent  are obtained in the far field approximation assuming a small bandwidth $\Delta \omega$, which is the frequency spectrum of the absorbed light. Here $\tau_c$ is the coherence time and $\nu_{k,l}=2\pi\vert \vec{r}_k-\vec{r}_l\vert/l_c$, where $l_c$ is the transverse coherence length. 
The coherence time is inversely proportional to the bandwidth ($\tau_c=2\pi/\Delta\omega$)\footnote{This a rough approximation but valid for any kind of radiation field. Given a known temporal coherence function $\Gamma(\tau)$, the coherence time is defined as $\tau_c \equiv \sqrt{\frac{\int^{\infty}_{-\infty} \tau^2\vert\Gamma(\tau)\vert^2}{\int^{\infty}_{-\infty} \vert\Gamma(\tau)\vert^2}}$ \cite{Mandel}.} and the transverse coherence length is inversely proportional to the square root of the solid angle covered by the source ($l_c=\lambda_0/\sqrt{\Delta\Omega}=\lambda_0 \frac{D}{a}$) with $D$ and $a$ being the distance to the light source and its diameter, respectively, and $\lambda_0$ is the mean light wavelength \cite{Mandel,Fox}. Because of this, the transverse coherence increases for larger distances from the source -- where the wavefronts tend to be plane -- such that the propagation vectors at different positions on the detection surface are less spread around the perpendicular direction. Therefore, spatio-temporal correlations (Equations \ref{fcoh} and 3) will be relevant  when $T < \tau_c$ and $A<A_c$ . \\

The photo-counting probability to jointly detect $n_1,n_2 \ldots n_N$ photons $P(n_1,n_2,...,n_N;T)$, in a time window $T$  for such multiple photo-detection -- i.e. the detection time; or to observe an inter-photon detection time interval $t$,  $f(t)$,  \cite{Bedard,Bures71,Zardecki}, are useful to describe respectively,  spatial and temporal correlations in the photodetection events. They are obtained by differentiation of the generating function  \cite{Van_Kampen,Rockower}: $P(n_1,n_2,...,n_N;T)=\left \{ \prod_{i=1}^N \frac{(-1)^{n_i}}{n_i!}\frac{\partial^{n_i}}{\partial s_i^{n_i}}  \right \}G(\{s_i\},T)|_{\{s_i=1\}}$ and $f(t)=-\left.\frac{dP(n=0,T)}{dT}\right|_{T=t}$. See section I of supplementary material (SM) for more details.\\

\section{Results and discussion: Effects of light correlations on light harvesting} \label{results}

\subsection{Spatial configuration effects} 

Spatial correlations for a detection time  $T \ll \tau_c$ are described by the generating function

\begin{equation}
G(\left\{s_i\right\},T)=\prod^{N}_{k=1} \left[ 1+\left( s_1s_2\ldots s_N \alpha_1\alpha_2\ldots \alpha_N)\right)^{1/N}  T \left\langle I_k \right\rangle b'_k \right]^{-1}
\label{G_Tpeq}
\end{equation}

\noindent where $b'_k$ are the  eigenvalues of the spatial Fredholm equation, normalized by $\langle I_k \rangle$, which  is the average intensity (photons per $T$ time) reaching the $k$-th detector, whose efficiency is $\alpha_k$. If all the detectors are identical, the total intensity gets $\langle I \rangle= N\langle I'\rangle$, where $\langle I' \rangle=\langle I_k \rangle \hspace{2mm} \forall k$. The total number of detected photons is $\langle n \rangle= (\alpha A) T N \langle I' \rangle$, and therefore, in cases where the $\alpha$, $A$, $N$ and $T$ are fixed, we will refer to $\langle n \rangle$ as \comillas{intensity} indistinctly.\\
		
\begin{figure}[ht]
			\centering
			\includegraphics[width=1.1\textwidth]{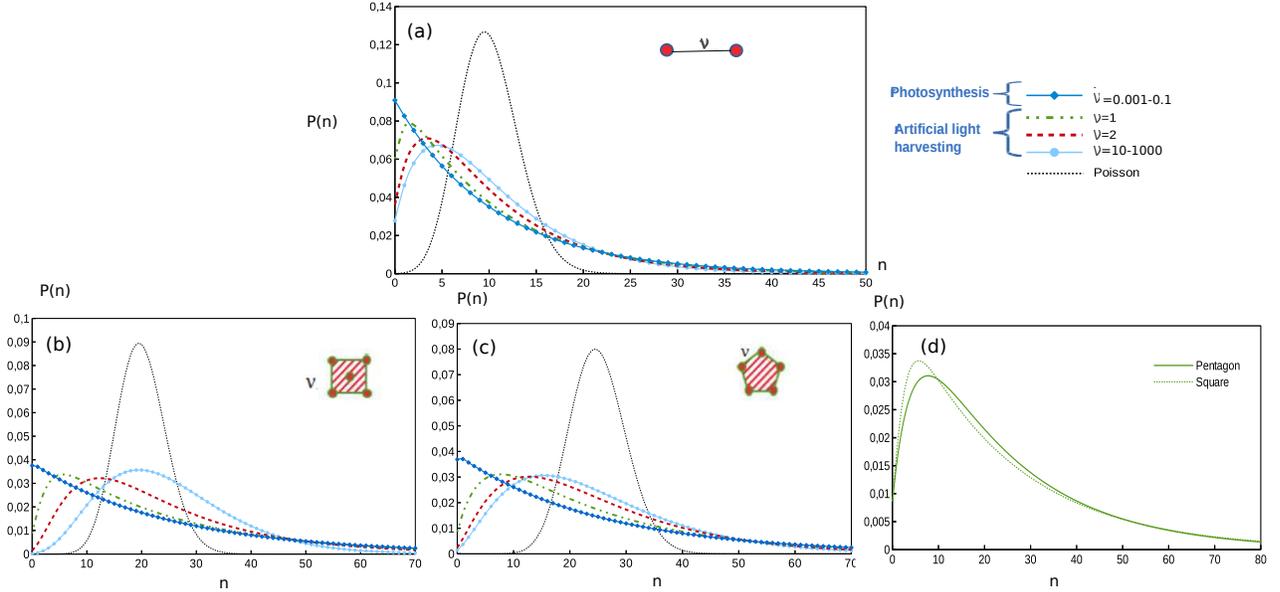}
		\caption{\footnotesize{$P(n)$ probability distributions for the total number $n$ of detected photons per detection time $T$, in a set of $N=2$ detectors in (a) and for a set of $N=5$ detectors in (b)-(d). In all the plots $\langle n \rangle/N=5$ photons per detector and $T/\tau_c\ll 1$.}}
		\label{Pnsimple}
	\end{figure}

Under these conditions, the photo-counting probability distributions for two detectors $N=2$, as a function of their normalized separation $\nu_{k,l}$ are shown in Figure \ref{Pnsimple} (a). For simplicity in the notation, we omit subscripts ($\nu_{k,l}=\nu$) if it is not required to specify the detectors addressed. It is important to remark that $\nu$ is defined as the ratio between the detectors separation and the coherence length, so that, values below 1 correspond to detectors within the coherence length. While for short separation ($\nu \leq 0.1$) the distribution peaks at low photo-counts numbers reminiscent of Bose Einstein statistics, for larger separations the peak shifts to higher number of events with a rather Poisson-like profile. Figure \ref{Pnsimple} (b) and (c) display the same calculations for a set of 5 detectors covering equal areas but organized differently. These situations present, for a separation $\nu\simeq 1$,  different photo-count distributions, as Figure \ref{Pnsimple} (d) corroborates. Similar results are obtained for $N=3$ and $N= 4$ detectors (cf. SM figures 2 and 3). This result implies that sunlight with transverse coherence length $l_c\simeq 50 \mu$m \cite{Mashaal2}, will imprint distinctive features in the photo-detected time traces from harvesting units with detectors set at about $l_c/2\pi\simeq 10 \mu$m apart.  Otherwise, detectors farther apart will show characteristics similar (but not equal to) uncorrelated events, just as Poissonian statistics forecasts.\\

\begin{figure}[ht] 
		\centering
			\includegraphics[width=1.1\textwidth]{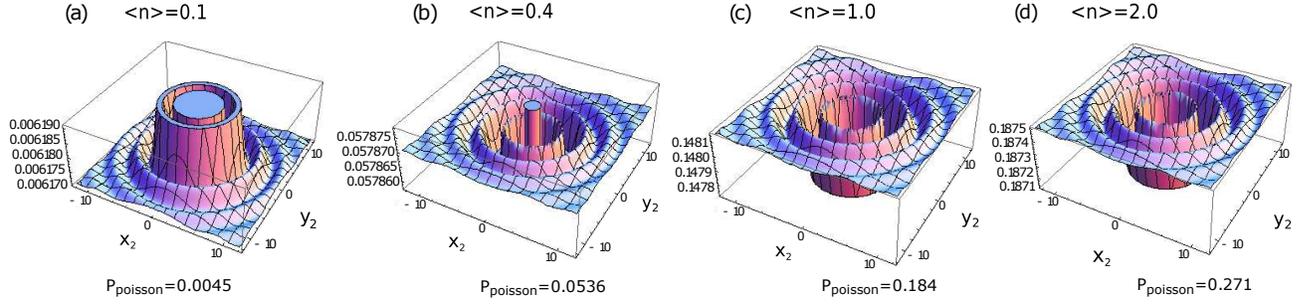}
		\caption{\footnotesize{$P(n=2)$ 3D plots for $N=2$ punctual detectors. The first one is placed at the origin and the second scans the 2D space ($x_2,y_2$), scaled in $\nu$ units. }}
		\label{Pn3DT2}
	\end{figure}

The resemblance of the Bose-Einstein distribution with  $P(n)$ for $\nu\ll 1$ and $T\ll \tau_c$, besides the trend at longer distances to exemplify uncorrelated events, illustrates that photo-detection  results from the characteristics of the light  (correlations and intensity), convoluted with the ability of the detector's geometry to reproduce these features. For this case, the probability $P(n=0)$ is maximal, and expresses the sparse long inter-photon times in between bunched packets.  When measuring correlated light within a coherence area $A_c\propto l_c^2$, the probability of the next absorption strike depends on the distance from the previous photo-detection. To underline the interplay between the detector's configuration and the field properties, Figure \ref{Pn3DT2} displays how the intensity and correlations of the incident field influences the joint probability density $P(n=2)$, at different positions of a second detector when the first detector is placed at the origin. Despite the changes in the probability values are small, it can be seen that there is an improvement at low intensities (Figures \ref{Pn3DT2} (a) and (b)), when comparing with the Poissonian probability values which are configuration independent (displayed  as a text inset for each $\langle n \rangle$). Whether an improvement for thermal light detection is available, depends on the light intensity; slightly greater intensities (Fig. 2(c)-(d)) show that Poissonian detection probability is higher within the detection area being scanned.\\

\begin{figure}
		\centering
			\includegraphics[width=1.1\textwidth]{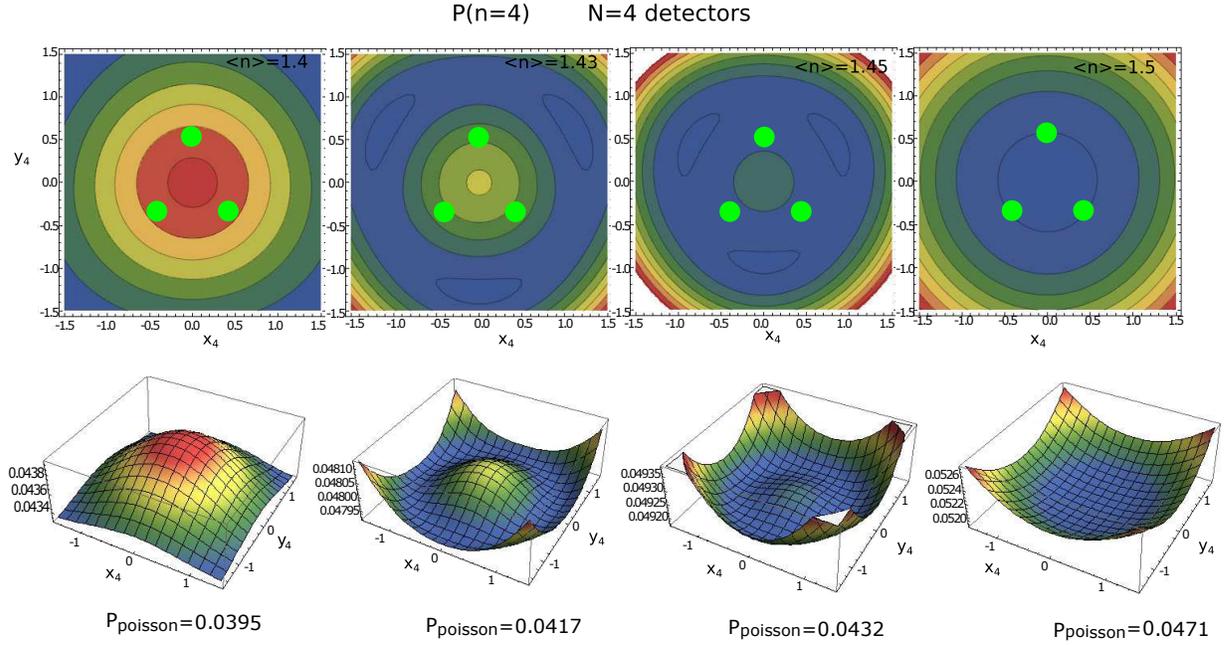}
		\caption{\footnotesize{$P(n=4)$ contour and 3D plots for configurations of $N=4$ detectors. Three detectors are fixed in the equilateral triangle vertices, the fourth one is scanned and its coordinates $(x_4,y_4)$ are in dimensionless $\nu$ units. The triangle's side is $\nu=1$. Detection time  $T/\tau_c\ll 1$. }}
		\label{Pn3DT4Tr}
	\end{figure}

Since on actual technologies detection devices can register finite numbers of events per counting time, we show the probabilities $P(n=n')$ to detect a fixed $n'$ number of photons per $T$ for different configurations and intensities. Figure  \ref{Pn3DT4Tr} displays the detection probability $P(n=4)$ for a set of $N=4$ detectors with three of them fixed at the vertices of an equilateral triangle of side $\nu=1$. Notice that for intensities equal to $\langle n \rangle= 1.4$, the detection probability shows a maximum for the placement of the fourth detector within the coherence area $A_c$ (cf. Figure\ref{Pn3DT4Tr} (a)), while for slightly higher  intensities (cf. \ref{Pn3DT4Tr} (b)-(c)) the detection patterns present a transition such that $P(n=N)$ is greater when the fourth detector is outside this area. As will be in short stated, for either lower intensities -which represent solar irradiance- or much greater irradiance, photodetection pattern are significantly different and may present several maxima within $A_c$. The  correlations present in thermal light improve the detection probability  by about 0.1-1\%, in comparison with the result obtained for Poissonian sources, provided in the text insets. \\

\begin{figure}[ht]
			\centering
			\includegraphics[scale=0.65]{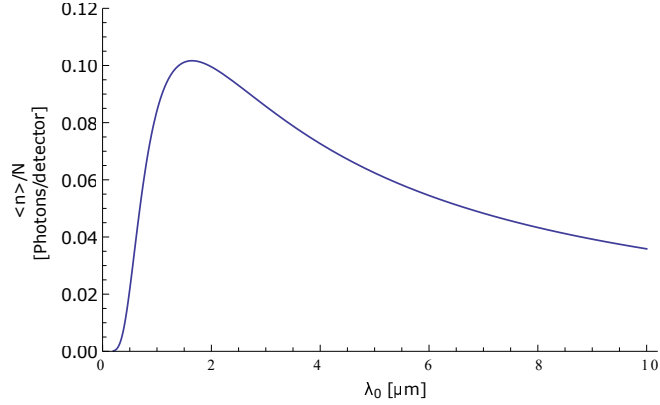}
		\caption{\footnotesize{Maximum average number of correlated photons $\langle n \rangle_{max}$ over the coherence time and area as a function of the light wavelength $\lambda_0$ for band width $\delta \lambda=100$ nm). Here $N=1$, $\alpha=1$, $A_C=\pi(50\mu$m$)^2$, $\tau_c=2\pi/\Delta \omega$.}}
		\label{<n>}
	\end{figure}

\begin{figure}
			\centering
			\includegraphics[width=\textwidth]{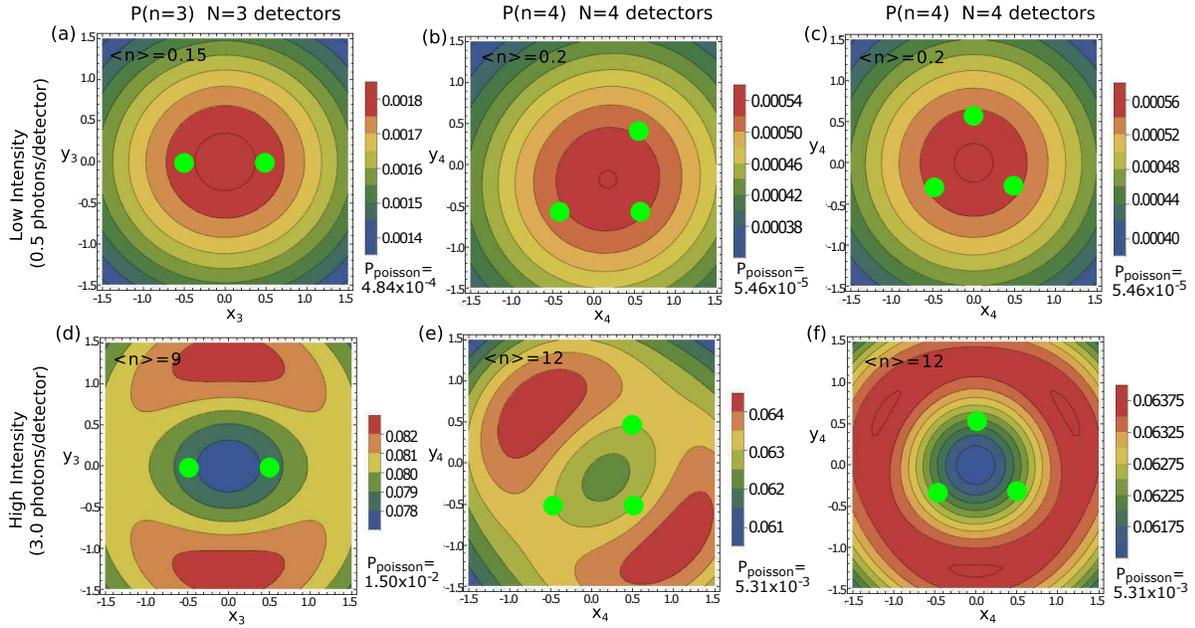}
		\caption{\footnotesize{(a) and (d) $P(n=3)$ for $N=3$ detectors. Two of them are fixed at shown positions and the position of the third one is scanned.  (b),(c), (e) and (f) $P(n=4)$, for $N=4$ detectors. Three are fixed at the shown positions and the position of the fourth is scanned. All the calculations made for $T/\tau_c\ll 1$ and the scanned coordinates $(x_i,y_i)$ in dimensionless $\nu$ units. (a)-(c) Low intensities $\langle n \rangle/N = 0.05 \leq 0.1$. (d)-(f) High intensity $\langle n \rangle/N =3 \gg 0.1$. }}
		\label{combined}
	\end{figure}

In order to explore the possibility of finding such increased probability of the central detector, it is important to underline the realistic intensities, transverse coherence area and coherence time for sunlight on Earth. Regarding the coherence area is $A_c =\pi l^{2}_c\approx 8 \times 10^{-9}$m$^2$, for sunlight on Earth\cite{Mashaal1,Mashaal2}). Regarding the coherence time $\tau_c$, it ranges between 1 to 5 fs for crystalline Silicon detecting light spectrum between 400-1200 nm ($\Delta \omega_{max}\approx3$PHz)\cite{Van}, or can even be engineered in solar antennas which are designed to absorb at a specific wavelength or at a broad spectral bandwidth \cite{Briones}\footnote{Notwithstanding, the coherence time $\tau_c$ will depend on the absorption spectrum of the specific technology, i.e., on factors such as the material's bandgap, layers, temperature, and impurities, among others.}. Realistic values for the average number of detected photons $\langle n \rangle=\alpha TAN\langle I'\rangle$, at which correlations can be perceived	 by the detection system, depend on the interplay between the solar flux of photons $\phi(\lambda)$ -which peaks at about 500-600 nm-, and the detectors response ($\alpha TA$). Therefore, the maximal number of correlated detection events that can be registered per detector is limited by the coherence area and time ($TA \leq \tau_c A_c$) 

\begin{equation}
 \frac{\langle n \rangle}{N} \leq \alpha \tau_c A_c \int^{\lambda_0+\delta\lambda}_{\lambda_0-\delta\lambda} d\lambda \phi(\lambda)
\end{equation}

\noindent with

\begin{eqnarray}
\tau_c =\frac{(\lambda_0-\delta\lambda)(\lambda_0+\delta\lambda)}{2c\delta\lambda} \hspace{1cm} \text{and} \hspace{1cm}
\phi(\lambda)= \frac{2\Omega c}{\lambda^4}\frac{1}{e^{hc/\lambda K_B \mathbb{T}}-1}
\end{eqnarray}

\noindent where $c$ is the speed of light in vacuum, $h$ is the Planck constant, $K_B$  is the Boltzmann constant and $\mathbb{T}=5250$ K is the temperature of the Sun. Notice that $\tau_c$ is always positive since $\lambda_0 \pm \delta \lambda$ are the lower and upper limits of the band width. The scanning of $\lambda_0$ 
sets a maximum number of correlated detections (per detector) under natural conditions of $\langle n \rangle/N \approx 0.1$  at $\lambda_0=1.5\mu$ m in the near infra-red, as shown in Figure \ref{<n>}. 
The number of maximal correlated absorptions per detector ($\langle n \rangle/N$) is not changing as a function of the band width $\delta\lambda$, because 
of the trade-off between the sun-light intensity (at $\lambda_0$ and $\delta\lambda$) and the response of the detectors. When the integrated photon-flux is increased by a larger $\delta\lambda$, the detection area and time need to be smaller in order perceive the light correlations, and therefore $\langle n \rangle/N$ depends only on $\lambda_0$. In this intensity regime ($\langle n \rangle/N< 0.1$), Figure \ref{combined} shows that the probabilities calculated including correlations are about one order of magnitude greater than the Poissonian probabilities. Additionally, at low intensity the detection probability is always maximized when detectors configurations are dense within $A_c$ (from comparison of three to four detectors), and follow a crystallographic arrangement (from comparison of equilateral and scalene triangles). \\

\begin{figure}
		\centering
			\includegraphics[width=1.1\textwidth]{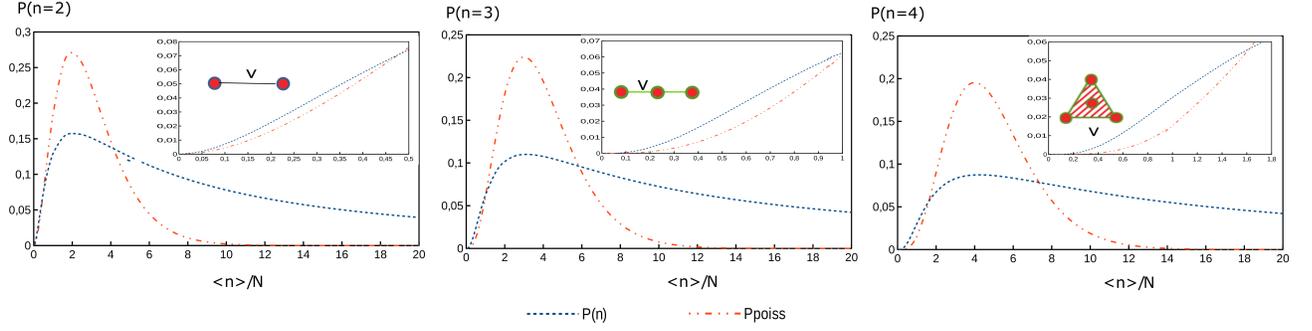}
		\caption{\footnotesize{Detection Probabilities $P(N=n)$ for $N=$2,3,4 detectors with  $T/\tau_c \ll 1$. The chosen configurations for N=3 and 4 detectors are the ones maximizing the $P(N=n)$ in Figure \ref{combined} (a) to (c). The insets show the corresponding Poissonian and correlated distributions in the low intensity regime.}}
		\label{PVsI}
	\end{figure}

Whether the configuration of detectors  can exploit the sun-light correlations to increase their photodetection rate for the natural intensity, can be explored by  the comparison of specific configurations subject to thermal or Poissonian light, presented in Figure \ref{PVsI}.  Here, we place the detectors at the position where the joint detection probability densities  from Fig.\ref{combined} is maximal, and check how $P(n=N)$ behaves - being $n$ exactly the number of detectors $N$- for different light intensities. In general, these probabilities differ importantly for thermal and Posionian light, but most importantly for the relevant intensity of sunlight reaching earth, the insets of Fig.\ref{PVsI} show that every configuration that exploits the sun-light correlations, enhances the probability to detect so many photons as the number of detectors present under the relevant intensity of sun-light. In fact, each detection device has a limited capacity to register arrivals per $T$ time; otherwise the average number of detection events would need to be integrated over the entire domain of $\langle n \rangle =\int_0^{\infty}nP(n)dn$ obtaining the same average as in the Poissonian case. Thereby, thermal light correlations can be exploited to increase the photodetection rate under the physical constraint of finite absorptions for detector's responsiveness.\\

\begin{figure}[ht]
			\centering
			\includegraphics[scale=1]{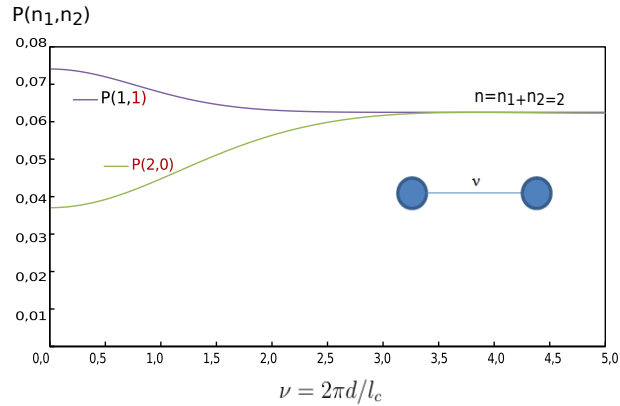}
		\caption{\footnotesize{Individual probability distributions. Two detectors separated by a normalized distance $\nu=2\pi d/l_c$ and the curves displays detection probabilities when distance is increased. In all the calculations $\langle n \rangle=1$ and  $T/\tau_c\ll 1$.}}
		\label{joint}
	\end{figure}

Another important consideration regards the detector's dead time, namely, the time after each event during which the detector is not able to record another event. For such technologies of detectors, it is preferable to have detection events distributed along the detection array and not concentrated in a single detector.  Figure \ref{joint} displays the probability to detect 2 photons in a two detector array, and shows that when the coherence of the sunlight is accounted for, if detectors are placed within the coherence length, the likelihood of consecutive detection at different detectors $P(1,1)$ is larger than that of detecting them in a single detector $P(2,0)$, and larger than that expected from uncorrelated events -when $\nu\gg 1$. The same results for a three detectors system are displayed in the SM (Fig.6), to corroborate that it is more likely to have detections distributed along the detectors set. 


\subsection{Temporal effects}

\begin{figure}[ht]		
	\centering
			\includegraphics[width=1.15\textwidth]{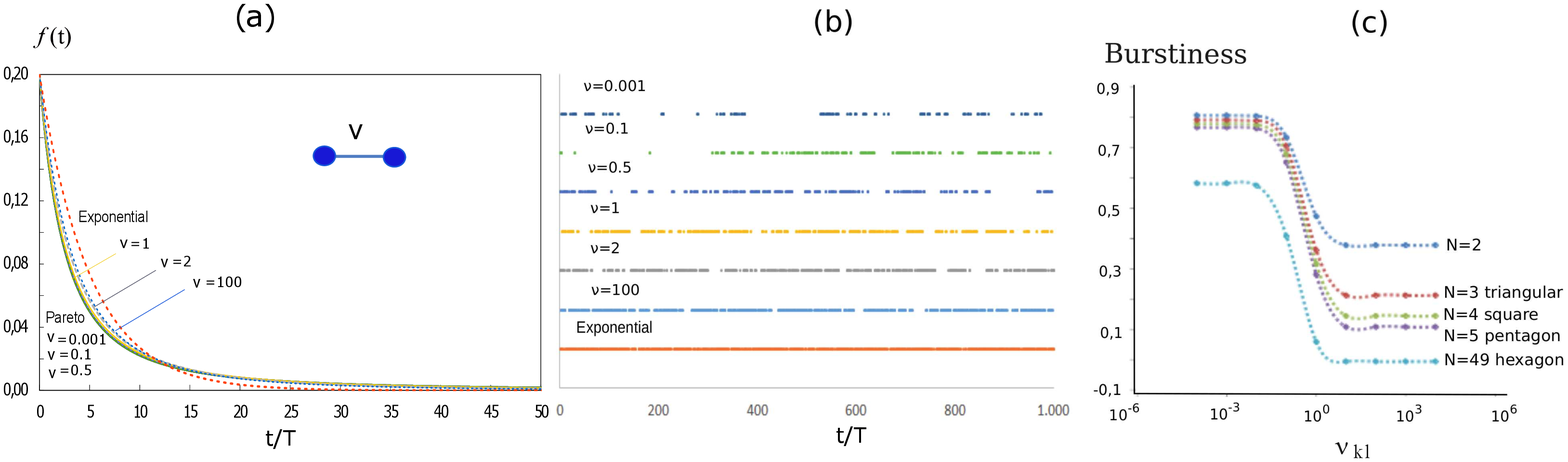}
		\caption{\footnotesize{(a) $f(t)$ density probability functions for $N=2$ detectors. 
(b) Corresponding simulated detections of photons.  (c) \textit{burstiness} as function of $\nu_{ij}$ (in semi-logarithmic scale) for $N=2,3,4,5,49$ detectors in different configurations. $\langle n \rangle/N=0.1$ and  $T/\tau_c\ll 1$ in all calculations.}}
		\label{fdt1}
	\end{figure}

Spatial and temporal correlations are complementary to produce clustered structure in the detection signals.  
In the Poissonian case, the number of detections per detection time $T$ is symmetrically distributed around a central value $\langle n \rangle$, but in correlated cases, $T$ intervals with few or no detection event are more likely and this is precisely what gives a clustered structure to the absorption events. Figure \ref{fdt1} (a) shows $f(t)$ for two detectors at different values of $\nu$. Here, longer-tailed distributions emerge when the distance is smaller than the coherence length, to reflect the polynomial decay of bunched thermal light.  Figure \ref{fdt1} (b) reflects that the apparently minor change in the distribution tails, impact severely the detected time traces because of the longer tails of the distributions. In order to better describe the effect of bunching, an adequate measure  is \textit{burstiness} \cite{Barabasi}, $B=\frac{\sigma-\mu}{\sigma+\mu}$, which quantifies the relation between the mean $\mu$ and the standard deviation $\sigma$. This measure should vanish, $B=0$, for a Poissonian process  and be positive (negative) for bunched (anti-bunched) statistics. Interestingly, the temporal correlations, due to its complementarity to the spatial correlation, are affected by the configuration of the array of detectors as displayed in Figure \ref{fdt1} (c). \\

The burstiness in the photodection can be used as a means to amplify the signal in technologies were single detectors are not able to produce enough electrons, useful for later stages, e.g. current rectification. For instance, rectennas (rectyfing antennas) technologies \cite{SAreview}, require an AC-DC rectifying stage achieved by diodes which require a current yield accomplished nowadays by the successful combination of the output from individual antennae. The bunched photon arrival, and hence several detections within a short time $T$, naturally amplify the signal for such technologies. Thus, besides the increment in the detection probability shown above, spatio-temporal correlations facilitate the currents rectification where the efficiency drop is important for this technology.\\

	\begin{figure}[ht]
		\centering
			\includegraphics[scale=0.50]{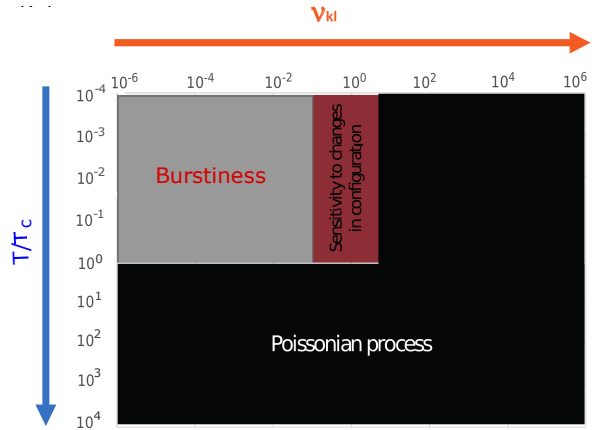}
		\caption{\footnotesize{Summary of the effects of spatio-temporal correlations at different detection times and separation scales. $\langle n \rangle=0.1$.}}
		\label{Tcrit}
	\end{figure}

Up to here we have accounted for $T\ll \tau_c$. 
Calculations of $P(n)$ and $f(t)$ are shown and detailed explained in the SM section 3, for different $T$ values. The obtained results confirm that correlations effects are robust against increment in $T$ up to the coherence time, and beyond this limit the detection statistics tend to be Poissonian, even when detectors are very near inside of a coherence area ($\nu_{k,l} \ll 1$ $\forall_{k,l}$).

\section{Conclusions}	


The present work outlines how the interplay between intensity and correlations of thermal light, is suggestive of densely packed and symmetric arrays of detectors, which shall be placed within the rather ample coherence area of sunlight, in order to enhance light harvesting technologies operative under normal sun-light conditions.  Moreover, the study of temporal correlations in the detection of thermal light in such designs, seems promising to overcome present limitations in the post-processing of electric currents.  Figure \ref{Tcrit} summarizes the regimes at which the complementary temporal and spatial correlations of sunlight can be useful, in order to highlight that spatio-temporal correlations of thermal light can be detected and then exploited, depending on the spatio-temporal scale of the detection; in fact, all the effects found for the maximal temporal correlations regime ($T\ll \tau_c$) are conserved when the detection time increases until the limit  $T=\tau_c$ where the statistics become Poissonian. This work opens, as a proof of principle, the perspective of exploiting specific geometries to enhance harvesting performance and facilitate the amplification-rectification of currents in present sun-light conversion technologies.

----------------
\section*{Acknowledgments}

A.de M. acknowledges partial financial support from COLCIENCIAS (Colombia) through the grant for doctoral studies in Colombia-call 528. A.de M., F.J.R. and L.Q. were partially supported by Banco de la Rep\'ublica de Colombia under grant 3646 (2016) and Faculty of Science - Universidad de Los Andes by projects 2014-2 and QUANTUM CONTROL OF NON-EQUILIBRIUM HYBRID SYSTEMS-Part II (2016). F.C.S. Acknowledges support from the EU project PAPETS, the ERC Synergy grant BioQ and the DFG via the SFB TR/21. N.F.J. is funded by National Science Foundation (NSF) grant CNS1522693 and Air Force (AFOSR) grant FA9550-16-1-0247.\\

\section*{References}


\providecommand{\newblock}{}

\end{document}


\title[Exploiting correlations in sunlight harvesting -Supplementary material]{Exploiting non-trivial spatio-temporal correlations of thermal radiation for sunlight harvesting -Supplementary material}

\author{A. M. De Mendoza$^1$, F. Caycedo-Soler$^2$, P. Manrique$^3$, L. Quiroga$^1$, F. J. Rodriguez$^1$, and N F. Johnson$^3$.}

\address{$^1$ Physics Department, Universidad de Los Andes, Cra 1 Nș 18A- 12 Bogot\'a, Colombia.}
\address{$^2$ Institut f\"ur Theoretische Physik, Universit\"at Ulm, Albert-Einstein-Allee 11, 89073 Ulm, Germany.}
\address{$^3$ Physics Department, University of Miami, Coral Gables, FL  33124 Miami, USA.}

\maketitle

\section{Thermal light statistics}	\label{Light}

Thermal light is partially coherent and statistically ranges between Poisson and Bose-Einstein distributions. Poissonian detection is the is the limit where light fields have no correlation in space and time, and which correspond to reception processes where the events are perfectly independent. On the other hand, the Bose-Einstein distribution describes the upper correlations limit which occurs for one mode thermal light. The Bose-Einstein probability distribution models the boson nature of photons, expressing their bunching tendency in such a way that arrival events become strongly correlated.\\

Below, we review the statistical behavior of the light with any degree of coherence by using the \textit{factorial moments generating functional} $G(\{s_i\})$ and afterwards we show how the limits are obtained therein. A generating functional is defined as an infinite series whose coefficients contain all the statistical information of a stochastic process. The probability distribution and the statistical cumulants of a given stochastic process are obtained by differentiation of the factorial moments generating functional with respect to the set of expansion parameters  $\left\{s_i\right\}$, as we shall see.
The factorial moments generating functional is defined \cite{Van_Kampen}:

\begin{equation}
G(\left\{s_i\right\})\equiv \exp \lbrace \sum_{m=1}^{\infty} \frac{(-1)^m(s_1s_2\ldots s_m)}{m!}k_{[m]}\rbrace ,
\label{Gs11}
\end{equation}

\noindent where $k_{[m]}$ is the m-th order factorial cumulant containing the m-th order correlation. The formula for $k_{[m]}$ can be seen as: 

\begin{equation}
k_{[m]}=(m-1)!(\alpha_1 \alpha_2\ldots \alpha_m) \Tr\left\{ \Gamma^{(m)}\right\}.
\end{equation}

\noindent In this expression, $\alpha_i$ is the efficiency or the effective cross section for absorption of the i-th detector and $ \Tr\left\{ \Gamma^{(m)}\right\} \equiv \sum^{N}_{l=1} \int^T_{0} dt \Gamma^{(m)}_{l,l}(t,t)$ is the trace operation, summing for all the $N$ detectors and integrating over detection time $T$\footnote{The observation time  $T$ in real scenarios is constrained by the time during which the detectors remain open to the incoming field.}. The $m$-th order correlation function is defined as \cite{Fox,Scully}:

\begin{equation}
\hspace{-1cm}  \Gamma^{(m)}_{l_1,\ldots,l_m}(t_1,\ldots,t_m)=\langle \hat{\vec{E}}^-(\vec{r}_{l_1},t_1)\ldots \hat{\vec{E}}^-(\vec{r}_{l_m},t_m)\hat{\vec{E}}^+(\vec{r}_{l_m},t_m)\ldots \hat{\vec{E}}^+(\vec{r}_{l_1},t_1) \rangle;
\end{equation}

\noindent where we assign an index $l_i$ for each detector, placed at the positon $\vec{r}_{l_i}$, and $t_i$ is the time at which the detector $l_i$ register the photon arrival. The $m$-th order correlation can be writen as \cite{Cantrell,Bedard}:

\begin{eqnarray}\label{ring}
\Gamma^{(m)}_{k,l}(t_1,t_2)&=\Gamma_{k,l}(t_1-t_2) \hspace{4.5cm}  \text{$m=1$, }\\
&=\sum^{N}_{i=1}\int^T_0 dt \Gamma^{(m-1)}_{k,i}(t_1,t)\Gamma_{i,l}(t,t_2) \hspace{1.2cm}  \text{$m\geq 2$ }
\end{eqnarray}

The definition presented in Eq.\ref{Gs11} is general since all the correlation orders of the stochastic process are included. However in light detection there is a set of $N$ detectors and only $N$-order correlations can be detected. Once the generating functional is defined, we can apply the formalism for thermal light detection which is Gaussian distributed and therefore just the second cumulant exists. After applying the equation in \ref{ring} with $m=2$, the thermal light generating functional in equation \ref{Gs11} turns into the double product $G(\left\{s_i\right\})=\prod_{j=1}^T \prod^{N}_{k=1} (1+\varpi_jb_ks_i)^{-1}$ presented in the main text (see Eq.1)\cite{Zardecki,Bures71}. Here $\varpi_j$ and $b_k$ accounts for the spatio-temporal correlations since they correspond to the eigenvalues of the temporal and spatial Fredholm equations presented there\footnote{Assuming cross spectral purity which separates the spatial from temporal correlations $\Gamma_{k,l}(t_1-t_2)=\sqrt{\left\langle n_k\right\rangle}\sqrt{\left\langle n_l\right\rangle}\gamma_{k,l}\gamma(t_1-t_2)$}.  The probability density function for the stochastic process is obtained by differentiation of the  generating functional. For light detection, the photo-counting probability to jointly detect $n_1,n_2 \ldots n_N$ photons in the detectors labeled 1,2,...,$N$ is given by:

\begin{equation}
P(n_1,n_2,...,n_N;T)=\left \{ \prod_{i=1}^N \frac{(-1)^{n_i}}{n_i!}\frac{\partial^{n_i}}{\partial s_i^{n_i}}  \right \}G(\{s_i\},T)|_{\{s_i=1\}}, 
\label{Eq:a1}
\end{equation}

\noindent and the probability of detection of $n$ photons in the total set, regardless of the specific counting record of any individual detector is
\begin{equation}
P(n)=\sum_{\{ n_i \}}\delta \left ( n-\sum_{i=1}^Nn_i\right )P(n_1,n_2,...,n_N) 
\label{Eq:a4}
\end{equation}

It is possible to map the photo-counting distribution $P(n)$ to the time domain in order to find the probability density function for the times between consecutive detections $f(t)$; it is obtained by time differentiation of $P(n=0)$ (probability to register zero arrivals when a time $T$ has passed) \cite{Van_Kampen,Bures71,Rockower}:\\

\begin{equation}
f(t) =-\left.\frac{dP(n=0,T)}{dT}\right|_{T=t};
\label{Eq:a2}
\end{equation}

Now we are going to apply the presented formalism to achieve the maximal and minimal correlation limits. 
The \textit{Poissonian limit} is reached when detection times and areas are so large that there are no spatial or temporal correlations.  This limit is obtained from Equation \ref{Gs11} with $m=1$ because only the first factorial cumulant exists for a non-correlated process and $\Gamma^{(1)}_{l,l}(t,t)=\left\langle n_l(t)\right\rangle$ \cite{Cantrell}. Therefore, the probability to detect $n$ photons is given by the Poisson distribution \cite{Fox,Scully,Mandel}:

\begin{equation}
  P(n)=\frac{e^{-\left\langle n\right\rangle}\left\langle n\right\rangle^n}{n!},
\end{equation}

\noindent where $\left\langle n\right\rangle=\sum^{N}_{l=1} \alpha_l \int^{T}_{0}  n_{l}(t)=NAT(\alpha \left\langle I'\right\rangle)$ is the average number of total detected photons in the detector's area $A$, during the counting time $T$. The net average intensity is $\alpha \left\langle I'\right\rangle$ for every detector with efficiency $\alpha$ such that both quantities ($\langle I'\rangle$ and $\alpha$) are position-independent. Thus, the Poisson limit is only reached when two conditions are fulfilled: first, no correlations exist because the area, time, and number of detectors are large ($NAT\rightarrow\infty$) and second, the intensity received by each one is very small ($\left\langle I'\right\rangle \rightarrow 0$) such that the total number of detections $NAT\left\langle I'\right\rangle$ is finite. For instance, even in the case of no spatial correlations, if the detection time is short enough, the statistics will not reach the Poissonian behavior, remaining correlated despite the distance among detectors. In the same way, a low intensity situation is not enough to guarantee Poissonian behavior and discard correlations. The interplay of these quantities is responsible of the finite correlations in the statistics associated to the photo-detected fields.\\

The \textit{Bose-Einstein limit} is obtained whenever the area covered by the set of detectors is much smaller than the coherence area ($A \ll A_C$) and  the detection time is also smaller than the coherence time ($T \ll \tau_C$). In this case, $\Gamma^{(m)}_{k,l}(t_1,t_2)=(NT)^{m-1}\left\langle I'\right\rangle^m$ no matter the positions of the detectors and the detection times. The differentiation of the obtained generating functional reproduces the Bose-Einstein distribution (eq. \ref{BE}), where the average detections at the $i$th detector $\left\langle n_i\right\rangle=\left\langle n'\right\rangle$ is also independent of position because of the assumption of a very small area. Again  $\left\langle n\right\rangle=NAT(\alpha \left\langle I'\right\rangle)$ is the average number of total detected photons in the detector's area $A$.

\begin{equation}
P(n)=\frac{1}{\left\langle n\right\rangle+1}\left(\frac{\left\langle n\right\rangle}{\left\langle n\right\rangle+1}\right)^n.
\label{BE}
\end{equation}


\section{Configuration sensitivity}	\label{Conf}

\begin{figure*}[ht]
		\centering
			\includegraphics[width=\textwidth]{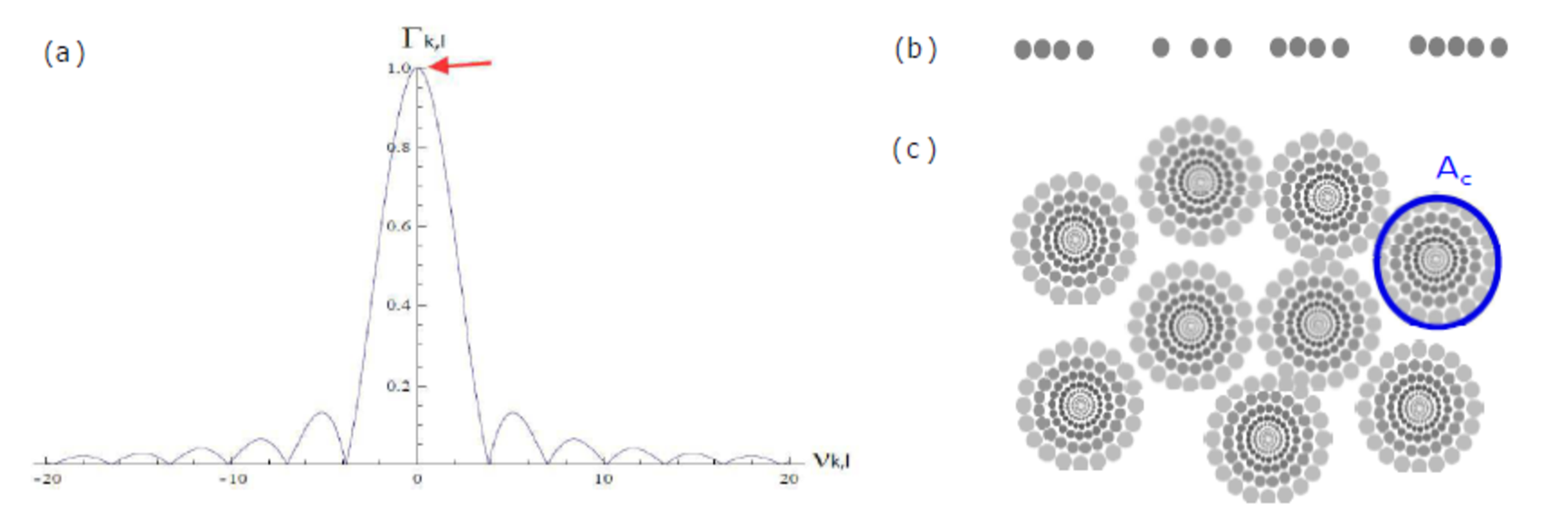}
		\caption{\footnotesize{Second order spatial-coherence function for the radiation field. The red arrow points the top of the function where first derivative tends to be constant. Scheme for the bunched photon-arrivals in time in (b) and for the clustering on the detection screen in (c) (instantaneous detection).}}
		\label{somb1}
	\end{figure*}

Sensitive and no-sensitive regimes are explained by the second order correlation function plotted in Figure \ref{somb1} (a). For very short separations (about $\nu < 0.1$), the top of the second order coherence function is flat and changes of $\nu$ about these value do not alter the spatial coherence, therefore configuration changes produce no effect on the photocounting probabilities. In the other extreme ($\nu > \pi$) the function  $\gamma_{ij}$ has fade out enough such that  spatial correlations are not relevant at those distances. In between of these values, ie. just before of the first zero of the Bessel function, the function's slop is steep enough to affect significantly $P(n)$.\\

\begin{figure*}
	\centering
		\includegraphics[width=\textwidth]{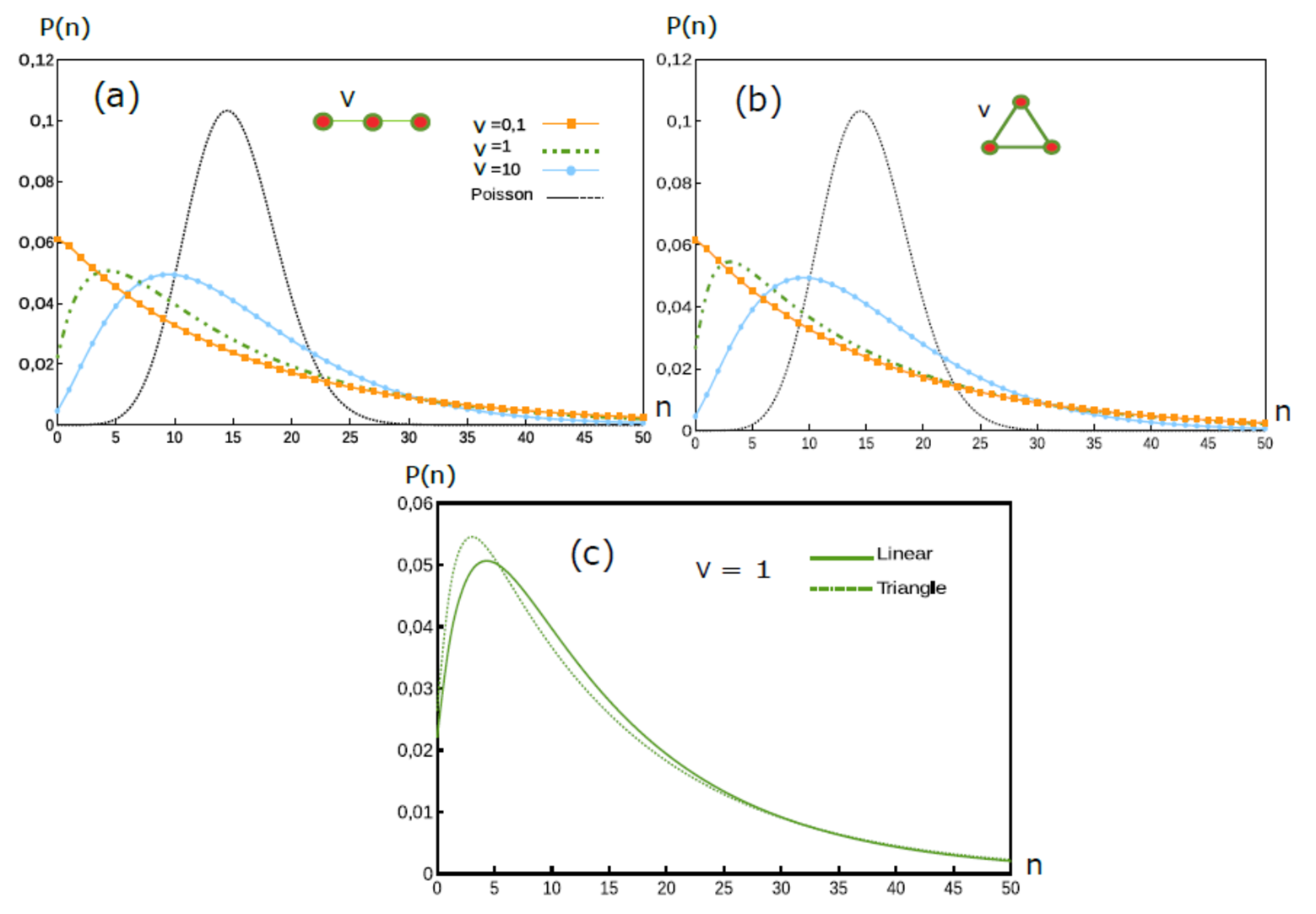}
	\caption{\footnotesize{Photocounting probability distribution $P(n)$ for $N=3$ detectors and different spatial correlation degrees $\nu=\nu_{k,l}=\frac{a}{r} \frac{2 \pi d_{k,l}}{\lambda}$. (a) Linear and (b) triangular configuration. (c) displays the comparison between the probability distributions of both configurations at $\nu=1$. $\langle n \rangle=10$ and  $T/\tau_c\ll 1$.}}	
	\label{tresdet}
\end{figure*}

\begin{figure*}[ht]
	\centering
		\includegraphics[width=0.9\textwidth]{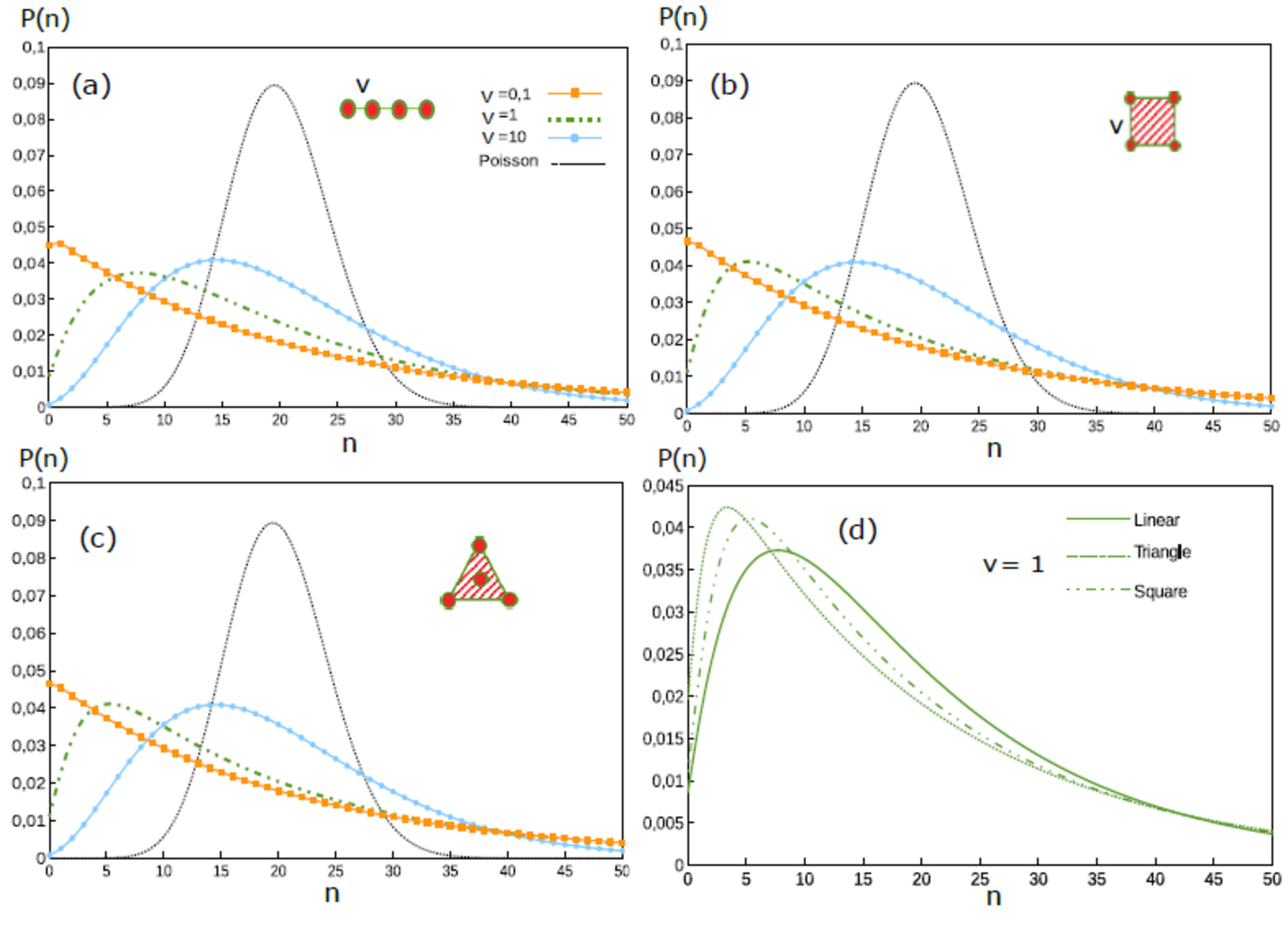}
	\caption{\footnotesize{Photocounting probability distribution $P(n)$ for $N=4$ detectors and different spatial correlation degrees $\nu=\nu_{k,l}=\frac{a}{r} \frac{2 \pi d_{k,l}}{\lambda}$. (a) Linear configuration. (b) Square configuration. (c) Triangular configuration with a detector in the middle, covering the same area as in (b).  (d) Displays the comparison between the probability distributions of configurations presented in (a) to (c) at $\nu=1$. $P(n)$ curves for $\nu\le 0.1$ are all overlapped and curves for $\nu\ge 2\pi$ as well, confirming that only for values of $\nu\approx1$ one has a sensitivity to configuration. $\langle n \rangle=10$ and $T/\tau_c\ll 1$.}}	
	\label{cuatrodeta}
\end{figure*}

Figures \ref{tresdet} and \ref{cuatrodeta} display the $P(n)$ curves for $N=3$ and $N=4$ detectors respectively, showing that configuration plays a role by changing photocounting probability distributions. In figure \ref{tresdet} (a) and (b), two configurations of three detectors (linear and triangular) are calculated and compared in (c). The $P(n)$ curves for $\nu\le 0.1$ are all overlapped and curves for $\nu\ge 2\pi$ as well, confirming that there is sensibility to changes in distances only when there is a competition between the detectors separation and the coherence length ($\nu_{ij}\approx 1$).
The comparison in Fig.\ref{tresdet}(c) shows that only for values of $\nu\approx1$ the distributions are different for different configurations. Similarly Fig. \ref{cuatrodeta} presents $P(n)$ probability distributions for three configurations of four detectors (linear, square and triangular) and compares them; leading to the same conclusion.\\
	
\begin{figure*}[ht]
			\centering
			\includegraphics[width=\textwidth]{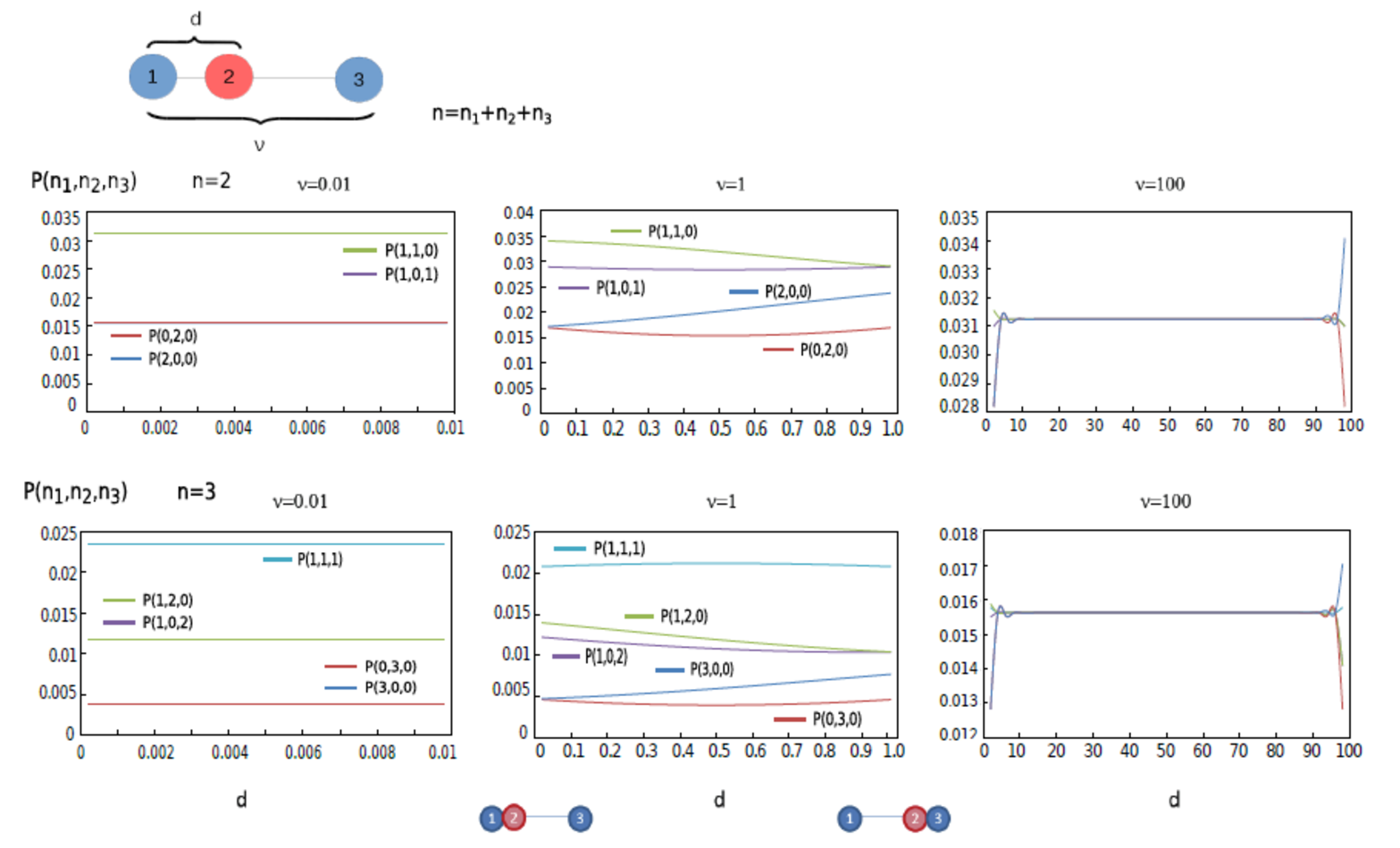}
		\caption{\footnotesize{Individual probability distributions (obtained from Eq.\ref{Eq:a1}) in a set of $N=3$ detectors. The positions of the extreme ones are fixed, separated by a distance $\nu$ and the middle one is mobile. Now, the distance between detectors labeled 1 and 2 is changed and the curves show detection probabilities in all the cases: $\nu=0.01$, $\nu=1$ and $\nu=100$. In all the calculations $\langle n \rangle=1$ and $T/\tau_c \ll 1$.}}
		\label{joint3}
	\end{figure*}

To figure out the reception behavior of the individual detectors, Figure \ref{joint3} displays the probability to detect 2 or 3 photons respectively in a set of $N=3$ detectors. In this example the detectors labeled 1 and 3 are at fixed positions separated a distance $\nu$ and the probability is calculated as a function of the position $d$ of the detector in the middle (labeled 2).  It is corroborated that in presence of correlations (plots for $\nu=0.01$ and $1$), the detection is more probable when photons are distributed among the detectors. When spatial correlations are not present ($\nu=100$) there are no preferred configurations \footnote{The curves are not flat in the extremes because when the mobile receptor is close to one of the other two, the correlations play a role again.}.  Specifically at $\nu=1$ and around, the probability to detect in one extreme increases when the mobile detectors is in the opposite side, confirming the result subsequently shown in Figs.4 and 5 at high intensity values.\\

\section{What happen when detection time increases?}\label{time-eff}

\begin{figure*}[ht]
			\centering
			\includegraphics[scale=0.55]{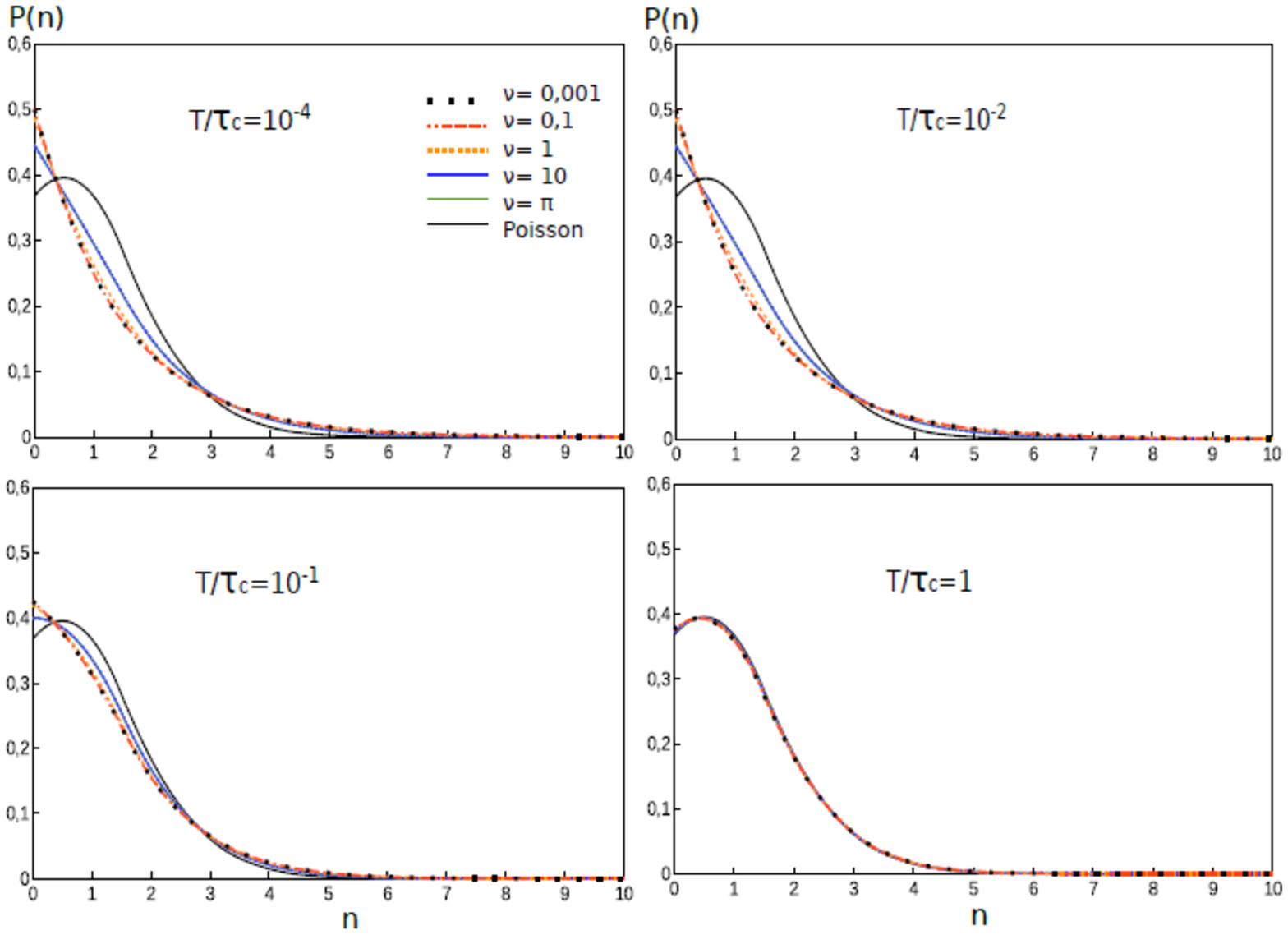}
		\caption{\footnotesize{$P(n)$ for N=2 detectors changing the detection time. (The average number of detections $\left\langle n\right\rangle=1$ is fixed).}}
		\label{PnTgr}
	\end{figure*}

As done for the short detection time limit, Figure \ref{PnTgr} explores the behavior of simple $P(n)$ curves for $N=2$ detectors when the detection time $T$ is increased. The plots show the approach to a Poissonian detection when $T\rightarrow \tau_c$. At this $T$ value, the sensitivity to configuration is almost lost and the $P(n)$ correlated curves practically overlap with the Poisson distribution. This sensitivity is found for the same $\nu$ range as in the short detection time case [$0.1 \le \nu  \le \pi$] but curves get closer and closer during the transition from maximally correlated to the Poisson distribution, when $T$ is increased.\\

\begin{figure*}
	\centering
		\includegraphics[width=1.05\textwidth]{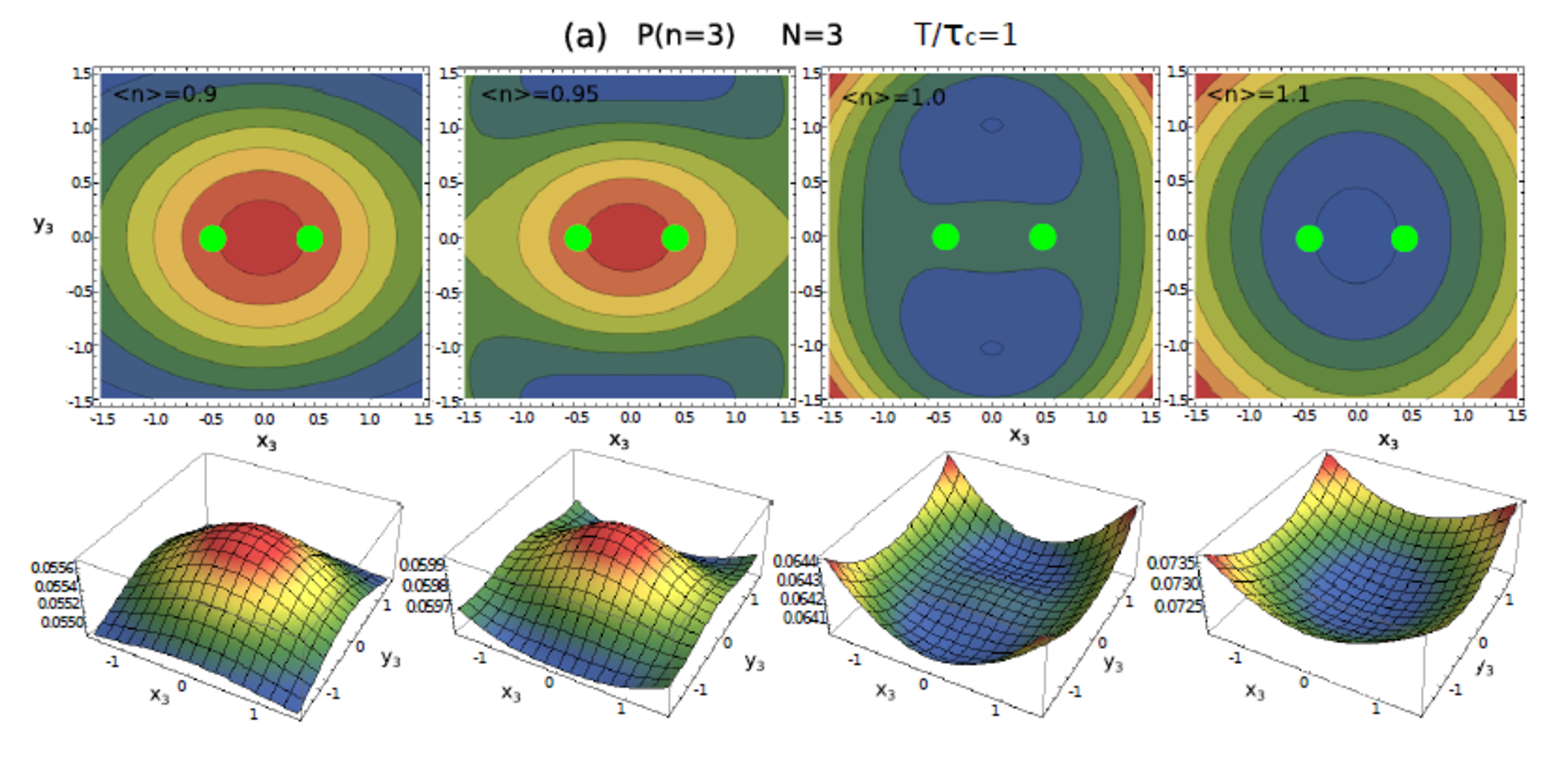}
	\caption{\footnotesize{$P(n=N)$ Contour and 3D plots for configurations of $N=3$. The detection time $T/\tau_c=1$. Two detectors are fixed separated a distance $\nu=1$ and the position of the third one is scanned. The $(x_3,y_3)$  coordinates are in dimensionless $\nu=\nu_{k,l}=2\pi r_{k,l}/l_c$ units.}}	
	\label{ContourTgr3}
\end{figure*}

\begin{figure*}
	\centering
		\includegraphics[width=1.05\textwidth]{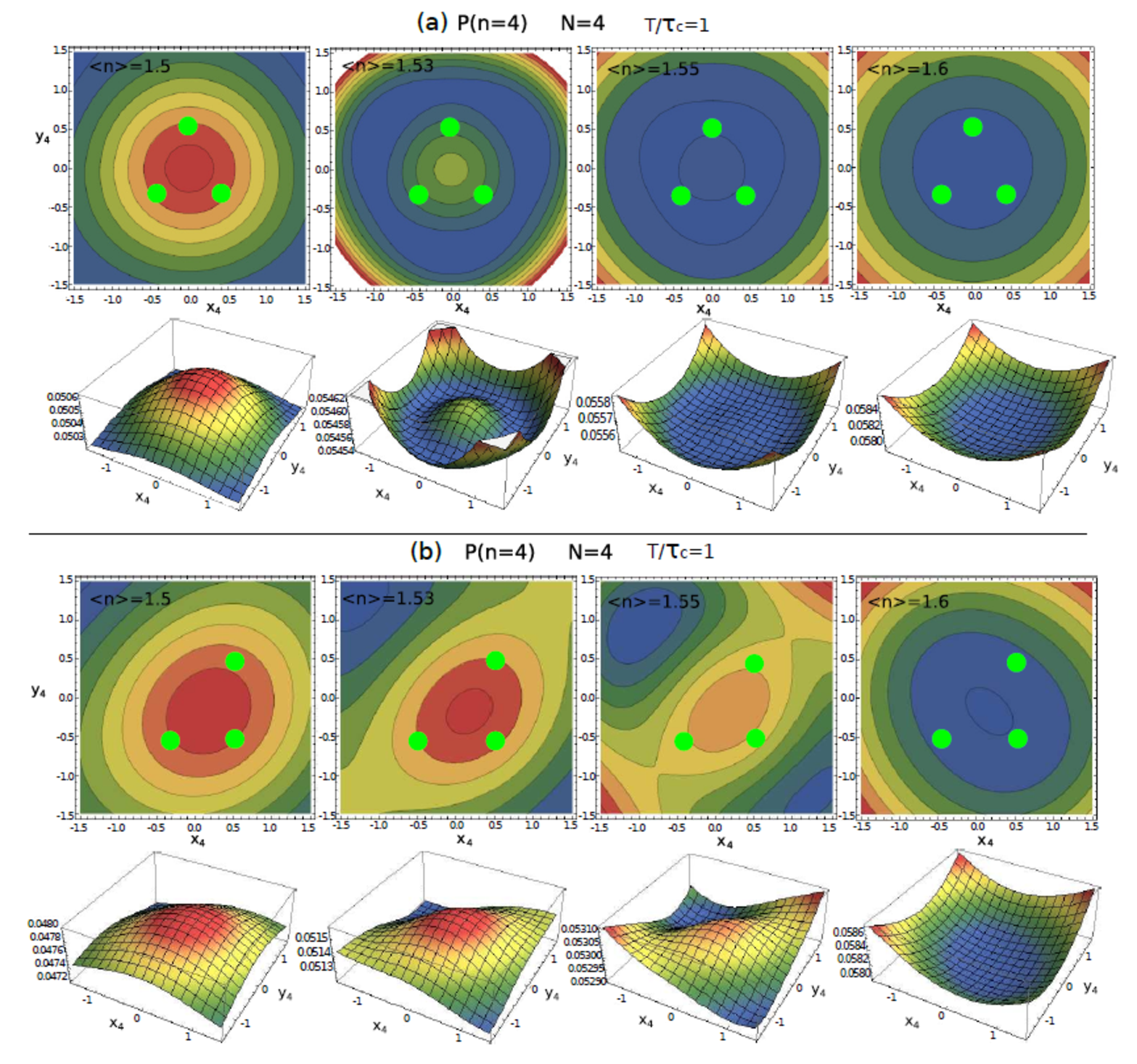}
	\caption{\footnotesize{Contour and 3D plots of $P(n=4)$ for configurations of $N=4$ detectors. The detection time $T/\tau_c=1$. (a) Three detectors are fixed in the vertices of an equilateral triangle and the position of the fourth one is scanned (The triangle's side is $\nu=1$). (b) Three detectors are fixed in the corners of square of side $\nu=1$ and the fourth one is mobile. The $(x_4,y_4)$  coordinates are in dimensionless $\nu=\nu_{k,l}=2\pi r_{k,l}/l_c$ units.}}	
	\label{ContourTgr4}
\end{figure*}

To better explore the photo-counting sensitivity to configurations around $\nu=1$, Figures \ref{ContourTgr3} and \ref{ContourTgr4} present contour and 3D probability distributions for sets of $N=3$ and 4 detectors respectively, for the detection time $T$ increased up to the coherence time. We found that the probability function displays the same behavior as for  $T\ll \tau_c$, but for slightly higher $\left\langle n \right\rangle$, the sensitivity to configuration is still conserved.\\

\begin{figure*}[ht]
			\centering
			\includegraphics[width=\textwidth]{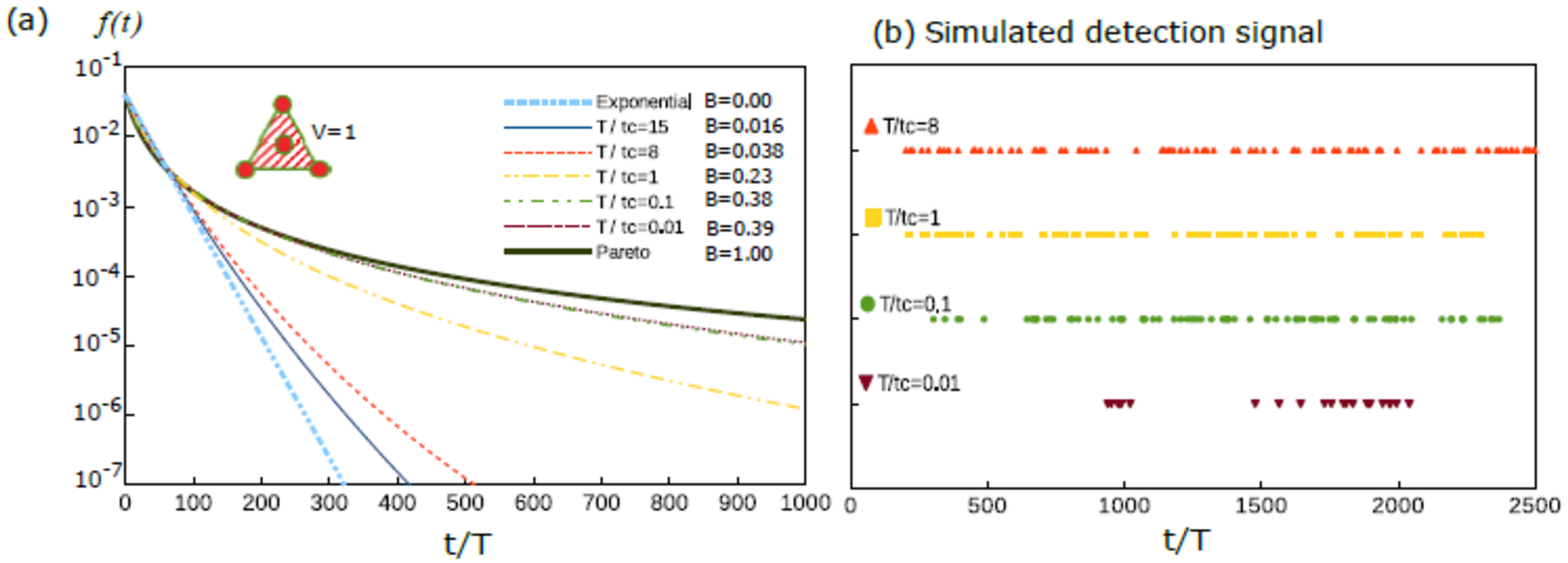}
		\caption{\footnotesize{(a) $f(t)$ probability functions for $N=4$ detectors in the shown configuration. $\langle n\rangle/N=0.01$ in a coherence time.  (b) Simulations of detection events, for some of the detection times in (a).}}
		\label{fdtTgr}
	\end{figure*}

Calculations of $f(t)$ for arbitrary detection times where also performed and are shown in figure \ref{fdtTgr}. Here the detection set is composed by 4 detectors in the configuration which maximizes the detection probabilities in Figures \ref{ContourTgr3} and \ref{ContourTgr4}. It can be seen how the correlation degree and therefore, the burstiness measure, decrease when detection time increases. This reduction confirms that for larger temporal counting window, the resolution of the measurement is less capable to detect the temporal correlations leading to more Poissonian behaviors.  Oppositely, the curves corroborates that for smaller detection times, $f(t)$ distributions become closer to the Pareto distribution (which corresponds to a Bose-Einstein process).  \\

\newpage

\providecommand{\newblock}{}